\documentclass[prl,letterpaper,twocolumn,tightenlines,superscriptaddress,showpacs,preprintnumbers,nofootinbib]{revtex4}

\usepackage{amsmath,amssymb,amsfonts,graphicx,pifont}

\begin{document}
\pagestyle{plain}

\preprint{VPI-IPNAS-10-06}

\title{Potential measurement of the weak mixing angle with
  neutrino-electron scattering at low energy}

\author{Sanjib Kumar Agarwalla}

\author{Patrick Huber} \affiliation{Department of Physics, Virginia
  Tech, Blacksburg, VA 24060, USA}

\date{\today}

\begin{abstract}

  We study the possibility to measure $\sin^2 \theta_W$ by
  neutrino-electron scattering at a value of the momentum transfer
  $Q\simeq30\,\mathrm{MeV}$ with a precision of $0.24\%$, which is
  only a factor three below the one obtained by LEP-I at the $Z$-pole.
  The neutrino source is a proton beam dump providing a clean beam
  from muon decay at rest and the detector is a $100\,\mathrm{kt}$
  scale water Cerenkov detector, which results in about 20 million
  signal events.
\end{abstract}

\maketitle

The Standard Model (SM) of particle physics provides a remarkably
accurate description of a wide range of phenomena in nuclear and
particle physics. The SM also unifies the weak and electromagnetic
forces into one gauge group, $SU(2)_L \times U(1)_Y$. The electro-weak
sector of the SM has been tested with utmost precision: in the weak
sector of the theory precisions at $0.1\%$ level are reached, whereas
in the electromagnetic sector of the theory the precision is 1 part
per billion or better. Despite its quantitative success, the SM has
been shown to be incomplete by the discovery of neutrino mass, the
existence of dark matter and the recent advent of dark energy. While
particle accelerators, like the LHC, continue to look for new physics
at ever higher energy scales, precision low energy observables have
been and continue to be an invaluable tool to learn about the scale of
new physics and to shed light into flavor sector, see {\it
  e.g.}~\cite{Erler:2004cx}. Precise tests of the electro-weak sector
of the SM are highly sensitive to the presence of oblique corrections
affecting vacuum polarization of the photon, $Z$ and $W$ bosons
through new particles in quantum loops and vertex
corrections~\cite{st}.  The weak mixing angle is defined by the ratio
of the $\mathrm{SU}(2)_L$ gauge coupling $g$ and the $U(1)_Y$ gauge
coupling $g'$ by following relation
$\sin^2\theta_W=\frac{g'^2}{g^2+g'^2}$. It is one of the key
parameters in this theory. Within the $\overline{MS}$ renormalization
scheme this is a scale dependent quantity and thus its value is
predicted to show a dependence on the energy scale at which it is
measured. This renormalization group running of the weak mixing angle
is an inevitable consequence of the electro-weak theory. Therefore,
the experimental demonstration of the running of the weak mixing angle
has been considered to be an {\it experimentum crucis} for the SM.

\begin{figure}[t]
  \includegraphics[width=\columnwidth]{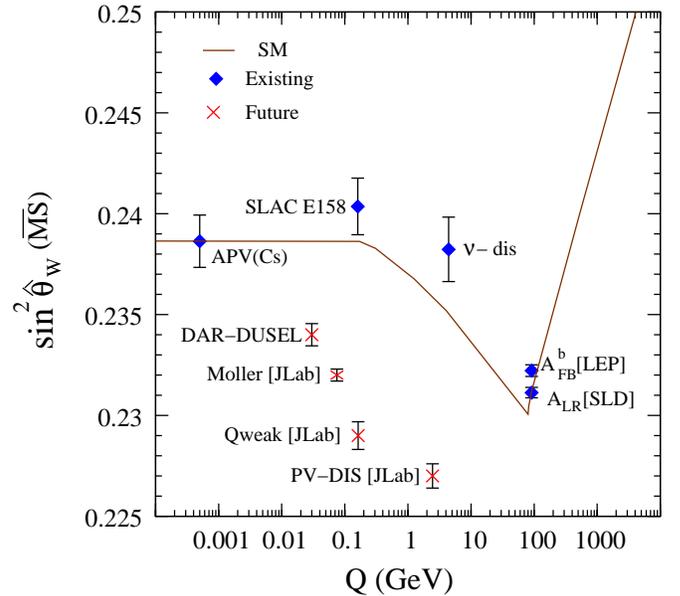}
  \caption{\label{fig:weinberg} A summary of the world data for the
    weak mixing angle as a function of the momentum transfer $Q$.  The
    solid curve depicts the running of $\sin^2\hat\theta_W$ in the
    $\overline{\mathrm{MS}}$ renormalization scheme~\cite{Erler:2004in}. The
    future experiments are shown with arbitrarily chosen vertical
    location. DAR-DUSEL denotes the result of this work.}
\end{figure}
The first measurement of the weak mixing angle was described
in~\cite{reines}, yielding $\sin^2 \theta_W = 0.29 \pm 0.05$. Modern
data on $\sin^2 \theta_W$ at low momentum transfer, $Q$, is based on
experiments on atomic parity violation at $Q \sim
10^{-5}\,\mathrm{GeV}$~\cite{Wood:1997zq} and SLAC E158, a M{\o}ller
scattering experiment at $Q = 0.16\,\mathrm{GeV}$~\cite{E158}.
Currently, the most precise measurement of $\sin^2 \theta_W$ was
obtained from $e^+e^-$ collisions at the $Z$-pole.  Interestingly, the
leptonic ($0.23113 \pm 0.00021$) and hadronic ($0.23222 \pm 0.00027$)
measurements of $\sin^2 \theta_W$ at $Z$-pole differ by 3.2 standard
deviations~\cite{lep}.  The NuTeV collaboration reported a 3$\sigma$
deviation from the SM value of $\sin^2 \theta_W$~\cite{nutev} using
deep inelastic neutrino-nucleus scattering. Many clever explanations
have been put forward to solve this puzzle~\cite{np}.  These
discrepancies could be a sign for new physics or maybe for not
understood experimental effects. To clarify the origin of these
discrepancies a set of new experiments to measure the weak mixing angle
at various values of $Q$ has been proposed and approved, these are
shown as red crosses in figure~\ref{fig:weinberg}.  There were several
other proposals to measure the weak mixing angle, not all of them shown
in figure~\ref{fig:weinberg}, using conventional muon neutrino
beams~\cite{cnb} at $Q \sim 0.1\,\mathrm{GeV}$ or electron neutrinos
produced from muon decay at rest~\cite{lsndnue}.  A 1 to 2\%
measurement of $\sin^2 \theta_W$ using a spallation neutron source was
proposed in~\cite{sns}. Another method was suggested in~\cite{reactor}
to measure $\sin^2 \theta_W$ to about 1\% at a reactor-based
experiment.  The possibility of measuring the weak mixing angle with a
precision of 10\% using low energy beta-beams was explored
in~\cite{ewbetabeam}.  At low $Q$, three new parity-violating electron
scattering experiments namely M{\o}ller scattering~\cite{Kumar:2009zz},
Q$_{weak}$~\cite{Pitt:2009zz}, and PVDIS~\cite{Souder:2008zz} 
have been proposed at Jefferson Laboratory and would provide 
precision measurements of $\sin^2 \theta_W$ in the range 
0.1 - 0.3\% in the absence of physics beyond the SM.

Neutrino-electron scattering is a simple, purely leptonic weak
interaction process that can play an essential role to prove the
validity and perform precision tests of the 
SM~\cite{Erler:2004cx,Marciano:2003eq} as well as many of its extensions. 
Neutrino electron scattering is also
quite sensitive to the weak mixing angle. In the SM, the 
tree-level\footnote{Higher order SM corrections are small and can be 
safely neglected.} differential cross section for neutrino-electron 
scattering can be written as~\cite{nuevogel}
\begin{equation}
\frac{{\rm d}\sigma}{{\rm d}T} = \frac{2G_F^2m_e}{\pi
E^2_{\nu}}\left[\alpha^2 E^2_{\nu}+\beta^2(E_{\nu} - T)^2
- \alpha\beta m_eT\right],
\label{eqn:diff_cross}
\end{equation}
where $G_F$ is the Fermi constant, $m_e$ is the electron mass, $E_\nu$
is the incoming neutrino energy, $T$ is the electron
recoil kinetic energy given by the kinematics and has the range $0 \le
T \le T^{\rm max} = \frac{E_{\nu}}{1+m_e/2E_{\nu}}$.  In the SM,
$\alpha$ and $\beta$ are process-dependent constants that depend on
$\sin^2 \theta_W$ as given in Table~\ref{tab:diff_cross}.  The
$\nu_\mu$-e cross section decreases with the increase in weak mixing
angle while $\nu_e$-e and  $\bar\nu_\mu$-e cross section increase.
\begin{table}[t]
\begin{center}
\begin{tabular}{|c||c|c|c|} \hline
   & $\nu_e e\rightarrow \nu_e e$ & $\nu_\mu e\rightarrow \nu_\mu e$ &
     $\bar\nu_\mu e\rightarrow \bar\nu_\mu e$\\ \hline
$\alpha$ & $\frac{1}{2} + \sin^2 \theta_W$ & $-\frac{1}{2} + \sin^2 \theta_W$ & 
$\sin^2 \theta_W$\\ \hline
$\beta$ & $\sin^2 \theta_W$ & $\sin^2 \theta_W$ & $-\frac{1}{2} + \sin^2 \theta_W$\\ \hline
\end{tabular}
\caption{\label{tab:diff_cross} The values of $\alpha$ \& $\beta$ in the SM 
for different processes involved in our case.}
\end{center}
\end{table}
The recoil electron is produced at an angle $\theta$ to the incident
neutrino direction where $\cos\theta =
(1+m_e/E_{\nu})/\sqrt{1+2m_e/T}$.  Thus the recoil electrons are
strongly peaked along the neutrino direction if $T \gg m_e$. This
forward peaking is used distinguish neutrino-electron scattering
events from neutrino reactions on nuclei in the energy regime from a
few MeV up to few tens of MeV. In this energy range, neutrino-electron
scattering can be observed experimentally in large water Cerenkov
detectors with high efficiency and with nearly no background, as
demonstrated by the results on solar neutrinos by
Super-Kamiokande~\cite{Fukuda:2008zn} and the Sudbury Neutrino
Observatory~\cite{Aharmim:2005gt}.

Here we suggest an experiment to measure $\sin^2 \theta_W$ up to a
precision of 0.24\% based on neutrinos from pion and anti-muon decay at rest
(DAR).  These anti-muons are obtained from the decay of stopped
$\pi^+$; the $\pi^+$ in turn are produced in a
proton beam dump. The novel concept compared to earlier proposals is
that we will use the energy dependence and \emph{not} the total rate,
{\it cf.}~\cite{lsndnue}, of the recoil electron spectrum which is
possible due to the much higher statistics.

Technologically, the key are high-intensity, low energy proton
sources, which currently are developed for various applications and
have been previously suggested as driver for a stopped pion neutrino
source~\cite{conrad}.  The electrons from neutrino electron scattering
will be observed in a large-size, $\mathcal{O}(100\,\mathrm{kt})$,
water Cerenkov detector like the one proposed for the Deep
Underground Science and Engineering Laboratory
(DUSEL)~\cite{dusel,duselwhite}.

\begin{figure*}[t]
\includegraphics[width=0.5\textwidth]{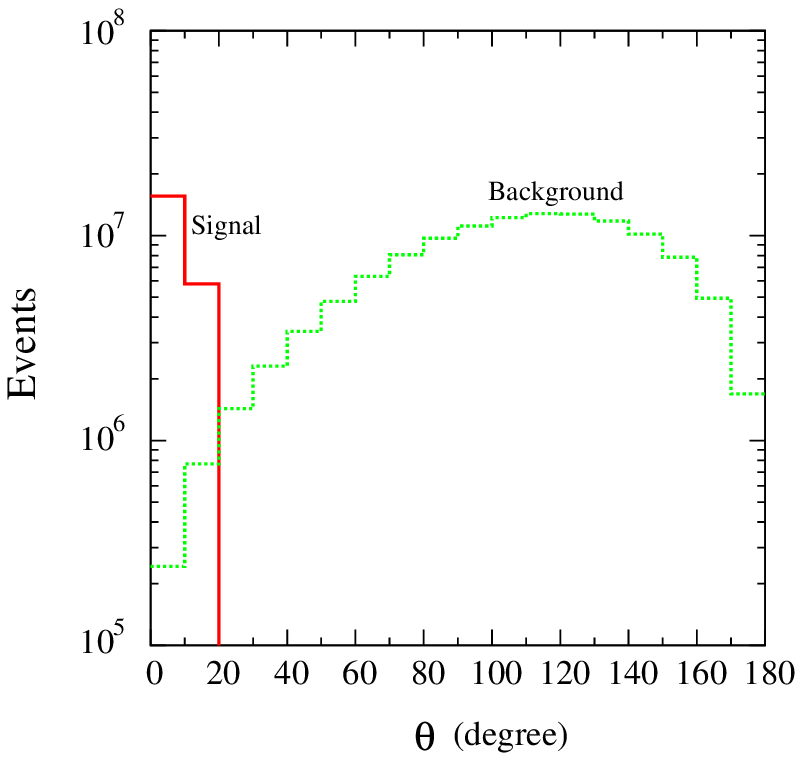}\includegraphics[width=0.5\textwidth]{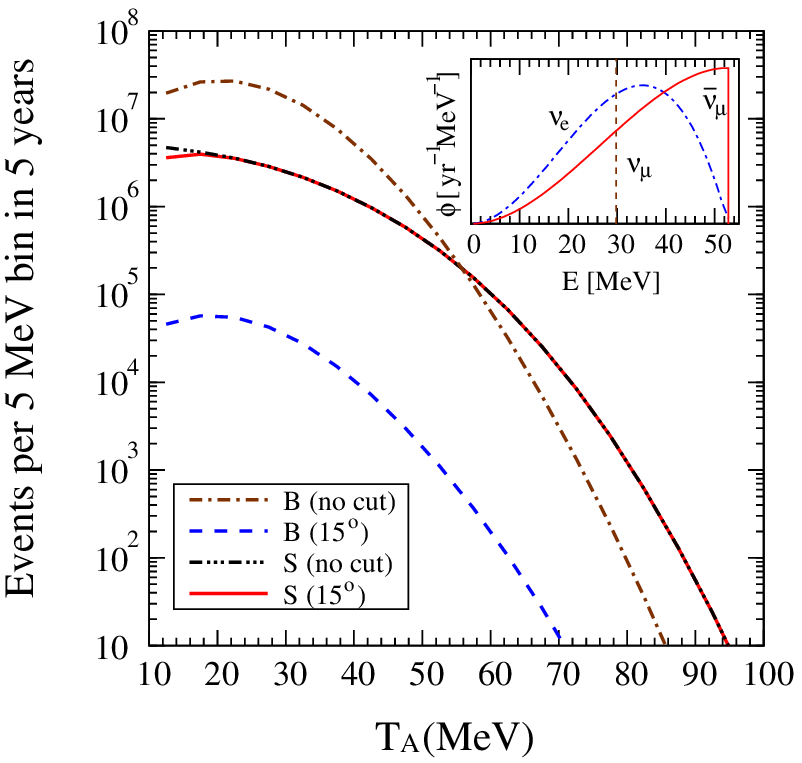}
\caption{\label{fig:events} The left hand panel shows the angular
  distribution of the signal and background events. The red solid line
  is the signal and the green dotted line indicates the background.
  The right hand panel shows the neutrino-electron scattering events
  in five years with two cyclotrons as a function of reconstructed
  recoil electron kinetic energy, $T_A$. The expected background
  events from charged-current $\nu_e$-Oxygen reaction are also shown.
  The impact of angular cut on signal and background events are also
  depicted. In the inset, the energy distribution of neutrinos in a
  DAR beam is shown.  }
\end{figure*}

We assume the same proton source parameters as in~\cite{conrad}:
$100\,\mu\mathrm{s}$ pulse every $500\,\mu\mathrm{s}$,
$2\,\mathrm{GeV}$ protons and $9.4\times10^{22}$ protons on target per
year, which results in $4\times10^{22}$ neutrinos per flavor and year.
These numbers apply for one of these sources, throughout of this paper
we assume 2 of them at the same location. The running time is 5 years,
thus the total exposure is $9.4\times10^{23}$ protons on target. The
resulting neutrino flux is shown in the inset of
figure~\ref{fig:events}. The flavor composition and spectrum are well
understood from first principles.  Due to the strong $\pi^-$
absorption in the target, the $\bar\nu_e$ contamination in the beam is
very tiny ($\sim 10^{-4}$) and we therefore will neglect it. For a
review on stopped pion neutrino sources and the associated physics,
see~\cite{spallation}.

We assume a $300\,\mathrm{kt}$ water Cerenkov detector consisting of
two volumes of right cylinder of $150\,\mathrm{kt}$ each, separated by
a distance of $60\,\mathrm{m}$. The neutrino source is in the middle
between the two detector modules so that both the detector volumes
will receive the same amount of neutrino flux.  Accounting for the
geometric acceptance, we find an average distance to target
corresponding to the actual distance between the source and the center
of mass of the detectors, which is $54\,\mathrm{m}$.

The incoming $\nu_e$, $\nu_\mu$ and $\bar\nu_\mu$ will scatter with
the electrons inside the detector and we will measure the kinetic
energy and the direction of the recoil electron. We assume 75\%
detection efficiency, $\epsilon$~\cite{Fukuda:2008zn}. The energy
resolution, $\delta T$, for the electron is $5\%\sqrt{\rm E/GeV}$
which is a conservative choice compared to
Super-Kamiokande~II~\cite{Fukuda:2008zn}. This corresponds $\delta
T=8.7\,\mathrm{MeV}$ at $T=30\,\mathrm{MeV}$.  We work with a bin
size of $5\,\mathrm{MeV}$ for $T_A$. We have checked that even if we
consider a bin size of $10-20\,\mathrm{MeV}$ the results will not
change appreciably.  We consider a reconstructed energy threshold of
$10\,\mathrm{MeV}$. Note, that the electron performance of a such
detectors can be calibrated to a very high level of precision using a
small electron linear accelerator~\cite{Nakahata:1998pz}, in
particular the energy scale can be determined to better than $1\%$.

\begin{figure}[t]
\includegraphics[width=0.5\textwidth]{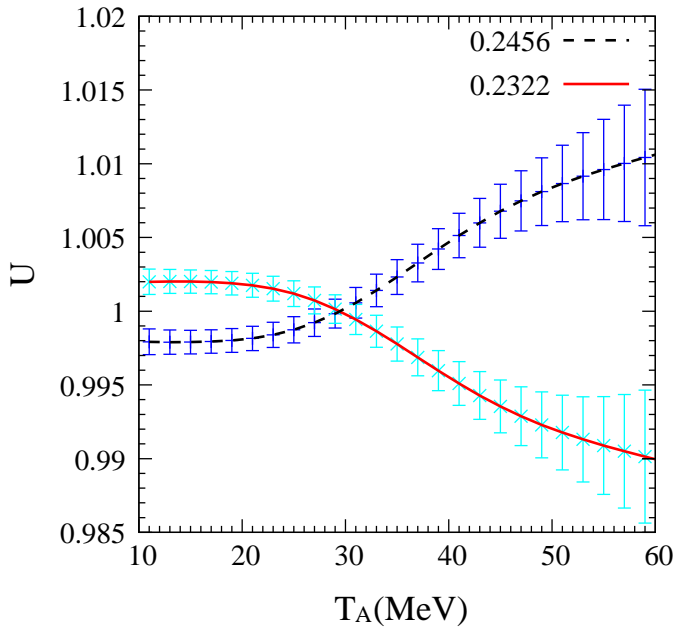}
\caption{\label{fig:shape} The variation of $U$ (as defined in the
  text) with $T_A$. The bin width is $2\,\mathrm{MeV}$ and the error
  bars are statistical $1\,\sigma$ errors.  The two lines are for
  different values of $\sin^2\theta_W$ as indicated in the legend.}
\end{figure}

The scattered electrons from the various neutrino flavors of a DAR source
are not distinguishable and their contributions must be 
incoherently added to calculate the sensitivity of $\sin^2 \theta_W$. 
The number of neutrino-electron elastic scattering events 
in the $i$-th bin of the reconstructed recoil electron kinetic energy, 
$T_A$ is given by
\begin{eqnarray}
{\rm N}_i &=&
\frac{c\,t\, n_e\,\epsilon}{4\pi}  \left\langle \frac{1}{ L^2}\right\rangle 
\sum_{{\rm flavors}}^{\nu = \nu_\mu, \nu_e, \bar\nu_\mu} \int_0^{E_{\nu}^{\rm max}} 
{\rm d}E_{\nu}\int_0^{T^{\rm max}}dT 
\nonumber \\ && 
\int_{T_{A_i}^{\rm min}}^{T_{A_i}^{\rm max}} {\rm d}T_A~R(T,T_A)
\phi_{\nu}(E_{\nu})
\frac{{\rm d}\sigma_{\nu}(E_{\nu},T)}{{\rm d}T},~~
\label{eqn:event}
\end{eqnarray}
where $c$ is the number of accelerators, $t$ is total running time,
$n_e$ is the number of target electrons in the detector, $\epsilon$ is
the detector efficiency and $R(T,T_A)$ is the Gaussian energy
resolution function of the detector.  $\langle 1/L^2\rangle$ is the
average inverse square distance of the detector from the source.
$\phi_{\nu}(E_{\nu})$ is the incoming neutrino flux and $E_{\nu}^{\rm
  max}=m_\mu/2$ is the maximum available energy of the neutrinos from
a DAR source, with $m_\mu$ being the muon mass. The resulting event
distribution as a function of $T_A$ is shown in
figure~\ref{fig:events} as solid line.

The by far dominant source of beam-on background are electrons
produced from charged-current reactions on oxygen, $\nu_{e} + {\rm O}
\rightarrow e^{-} + {\rm F}$~\cite{haxton}.  The angular distribution
of these background events is strongly backward peaked whereas the
neutrino-electron scattering signal events are sharply forward peaked,
as shown in the left hand panel of figure~\ref{fig:events}, where the
shape and size of these backgrounds is taken from Ref.~\cite{haxton}.
By applying suitable angular cuts, we can increase the signal over
background ratio~\cite{haxton,lsndnue}. The angular resolution of a
water Cerenkov detectors averaged over the energy range considered
here is better than $15^\circ$~\cite{Nakahata:1998pz}.  Beam-off
backgrounds are negligible in our case due to the deep underground
location~\cite{conrad}. The effect of the angular cut on the
background is shown in the right hand panel of
figure~\ref{fig:events}: the dash dotted line is the background
without any cuts, whereas the dashed lined is the background which
remains with a $15^\circ$ cut. As can be seen from
table~\ref{tab:result}, even a $30^\circ$ cut would be sufficient. The
importance of a tight angular cut is to eliminate the need to know the
angular distribution of the background to a high level of precision.

\begin{table}[t]
\begin{center}
  \begin{tabular}{||c||c||c||c||c||c||} \hline \hline
    $\theta$ & S & B & S/B & rel. error\\
    & & & & on $\sin^2\theta_W$\\
    \hline
    no cut & $21.2 \times 10^6$ & $122\times10^6$ & 0.17 & 0.57\% \\
    \hline
    30$^{\circ}$ &  $21.2 \times 10^6$ & $1.4\times10^6$ & 15 & 0.25\% \\
    \hline
    15$^{\circ}$ &  $19.8 \times 10^6$ &  $0.26\times10^6$ & 78 & 0.24\% \\
    \hline \hline
\end{tabular}
\caption{\label{tab:result} The expected number of signal and
  background events with and without angular cut have been given in
  second and third column respectively. The relative, $1\,\sigma$,
  error in measuring $\sin^2 \theta_W$ is quoted in the last column.}
\end{center}
\end{table}

To illustrate the effect, we are trying to measure, we define a shape
parameter $U$
\begin{equation}
U = 
\frac{ N_i(\sin^2 \theta_W)}{ \hat N_i(\sin^2 \hat\theta_W)}
\frac{\sum_{i=1}^n \hat N_i(\sin^2 \hat\theta_W)}{\sum_{i=1}^n{ N}_i(\sin^2 \theta_W)}.
\label{eqn:shape}
\end{equation}
where $N_i$ is the number of recoil electrons of the
reconstructed-energy bin $i$ as a function of the fitted value of
$\theta_W$, whereas $\hat N_i$ is the number of events predicted for a
given, true value of $\theta_W=\hat\theta_W$. $U$ obviously is
invariant under a rescaling of the total rate, therefore it contains
shape information only and $U\equiv 1$ for all energies if
$\theta_W=\hat\theta_W$. From the energy dependence of $U$, as shown
in figure~\ref{fig:shape}, we can extract the value of
$\sin^2\theta_W$.  The lines are for two different values of $\sin^2
\theta_W$.  In our analysis, we have taken the true value of weak
mixing angle $\sin^2 \hat\theta_W = 0.23863$, which corresponds to the
value measured at the Z-pole~\cite{pdg} evolved down to
$Q=0.03\,\mathrm{GeV}$ using the $\overline{MS}$ scheme.  The energy
dependence of $U$ on $T_A$ is due to the superposition of the recoils
from the mono-energetic $\nu_\mu$ line from $\pi$-decay and the
continuous spectra resulting from $\mu$-decay. This is also the reason
for the inflection point at $30\,\mathrm{MeV}$, which is the position
of the $\nu_\mu$ line. Since we are relying on the energy dependence
of our signal in order to suppress the impact of systematical errors,
we show in table~\ref{tab:precision} to what extent our results change
if the the energy resolution of the detector is varied.
%
\begin{table}[h]
\begin{center}
  \begin{tabular}{||c||c||c||} \hline \hline
    $\delta T$ & rel. error on $\sin^2\theta_W$\\
    ($\times \sqrt{\rm E/GeV}$) & ($1\,\sigma$)\\
    \hline
    5\% & 0.25\% \\
    \hline
    7.5\% & 0.29\% \\
    \hline
    10\% & 0.34\% \\
    \hline
    12.5\% & 0.4\% \\
    \hline
    15\% & 0.48\% \\ 
    \hline \hline
\end{tabular}
\caption{\label{tab:precision} The relative $1\,\sigma$ error in the measurement
of $\sin^2\theta_W$ as a function of the energy resolution of the detector.}
\end{center}
\end{table}
%
The precision on the weak mixing angle deteriorates by a factor of two
for a decrease in resolution by a factor of three. This in turn
translates into a requirement for sufficient photo cathode coverage
similar to the one of Super-K.

For our statistical analysis we use the so called pull approach as
used in~\cite{Huber:2002mx}. We bin our data in $T_A$ into bins of
$5\,\mathrm{MeV}$ and do not use $U$ but the event rates. We include a
$5\%$ systematic on the total number of signal events and a
(uncorrelated) $5\%$ systematic on the total number of background
events. Note, that systematics twice as large would not change our
results appreciably, demonstrating that this is a shape based
measurements.

\begin{figure}[t]
\includegraphics[width=0.5\textwidth]{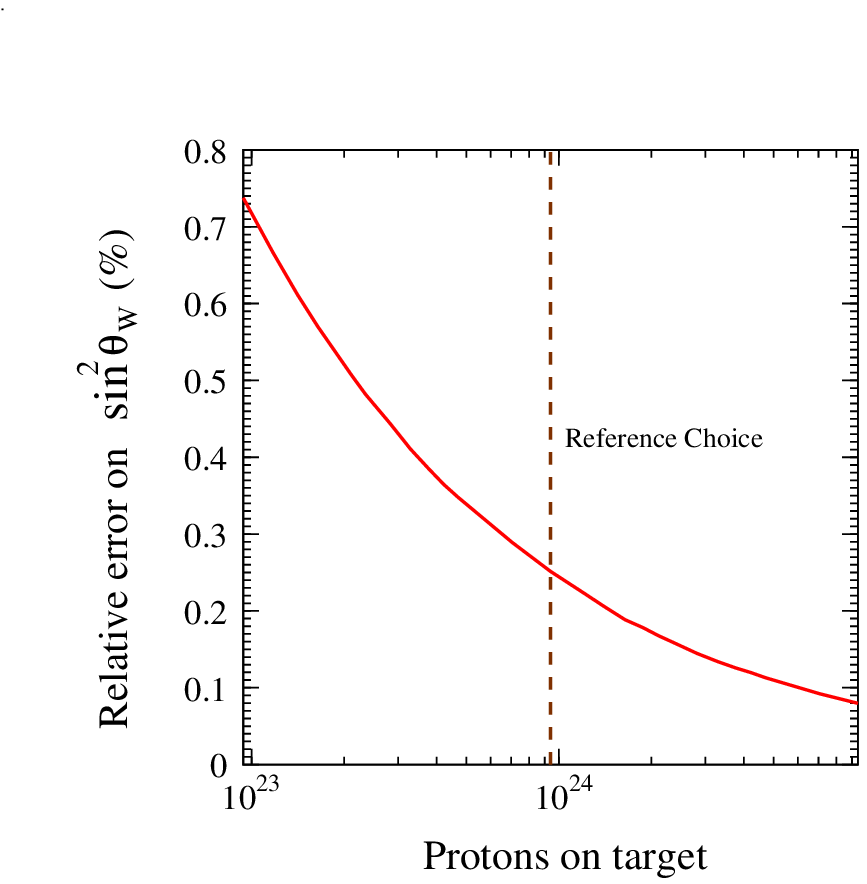}
\caption{\label{fig:exposure} The relative $1\,\sigma$ error on
  $\sin^2\theta_W$ in per cent as a function of exposure.  The dotted
  vertical line shows our reference choice for the exposure.}
\end{figure}

The expected number of signal and background events with and without
angular cut have been given the in second and third column of
table~\ref{tab:result}, respectively. The final precision of our
set-up in measuring $\sin^2 \theta_W$ is quoted in the last column. We
can see an improvement by factor of two when we impose an angular cut
of 30$^{\circ}$ on $\theta$ which helps us to get rid of the
background by a substantial amount. If we were to perform a rate-only
measurement instead, the precision in measuring $\sin^2 \theta_W$
deteriorates to 2.8\%, {\it i.e.} by one order of magnitude. The use
of the shape information is very effective in suppressing the impact
of systematics and thus, the question arises whether the measurement
is statistics limit with our current choice of beam luminosity and
detector size. Therefore, we study the impact of changing the exposure
over two orders of magnitude; the result thereof is shown in
figure~\ref{fig:exposure}. The error on the weak mixing angle scales
as $1/\sqrt{\mathcal{L}}$, where $\mathcal{L}$ is the exposure in
protons on target. This supports the conclusion that this measurement
is indeed statistics limited over a wide range of exposures.
Figure~\ref{fig:exposure} also allows to extrapolate the attainable
precision for different source-detector configurations. Note, that
when the detector mass (and hence the detector geoemetry) or
source-detector distance changes, the geometrical acceptance has to be
reevaluated, however in most circumstances the resulting correction is
expected to be relatively minor.

To summarize, the use of shape information, which we propose here for
the first time in this context, is the key to a high precision
measurement of $\sin^2\theta_W$ using neutrino-electron scattering.
The use of shape information is only possible if a sufficiently large
event sample can be obtained. To this end the combination of
high-intensity proton sources with very large detectors is crucial.
This configuration can be a natural part of the proposed physics
program for DUSEL. Our proposed experiment will provide a $\simeq
0.24\%$ measurement of $\sin^2 \theta_W$ comparable to the currently
most precise results.

We would like to thank J.M.~Link, M.~Pitt, R.~Raghavan, K.~Schollberg,
and, T.~Takeuchi for useful discussions. This work has been in part
supported by the U.S. Department of Energy under award number
DE-SC0003915.


\end{document}